\documentclass[aps,twocolumn,superscriptaddress]{revtex4}
\usepackage{graphicx}

\begin{document}

\title{Two-level systems driven by large-amplitude fields}

\

\author{S. Ashhab}
\affiliation{Frontier Research System, The Institute of Physical
and Chemical Research (RIKEN), Wako-shi, Saitama 351-0198, Japan}

\author{J. R. Johansson}
\affiliation{Frontier Research System, The Institute of Physical
and Chemical Research (RIKEN), Wako-shi, Saitama 351-0198, Japan}

\author{A. M. Zagoskin}
\affiliation{Frontier Research System, The Institute of Physical
and Chemical Research (RIKEN), Wako-shi, Saitama 351-0198, Japan}
\affiliation{Department of Physics and Astronomy, The University
of British Columbia, Vancouver, B.C., V6T 1Z1, Canada}

\author{Franco Nori}
\affiliation{Frontier Research System, The Institute of Physical
and Chemical Research (RIKEN), Wako-shi, Saitama 351-0198, Japan}
\affiliation{Physics Department and Michigan Center for
Theoretical Physics, The University of Michigan, Ann Arbor,
Michigan 48109-1040, USA}

\date{\today}

\begin{abstract}

We analyze the dynamics of a two-level system subject to driving
by large-amplitude external fields, focusing on the resonance
properties in the case of driving around the region of avoided
level crossing. In particular, we consider three main questions
that characterize resonance dynamics: (1) the resonance condition,
(2) the frequency of the resulting oscillations on resonance and
(3) the width of the resonance. We identify the regions of
validity of different approximations. In a large region of the
parameter space, we use a geometric picture in order to obtain
both a simple understanding of the dynamics and quantitative
results. The geometric approach is obtained by dividing the
evolution into discrete time steps, with each time step described
by either a phase shift on the basis states or a coherent mixing
process corresponding to a Landau-Zener crossing. We compare the
results of the geometric picture with those of a rotating-wave
approximation. We also comment briefly on the prospects of
employing strong driving as a useful tool to manipulate two-level
systems.

\end{abstract}


\maketitle

\newpage

\section{Introduction}

Two-level systems are ubiquitous in various fields of physics. A
large number of quantum phenomena rely on the existence of two
quantum states, or their underlying principles can be understood
using the simple model of a two-level system. Recently, two-level
systems have gained renewed attention as they represent the
building blocks for quantum information processing (QIP)
applications \cite{Nielsen}.

In the study of two-level systems, as well as many other quantum
systems, avoided level crossings are associated with a wide
variety of interesting phenomena. Large amounts of literature have
been devoted to analyzing the dynamics of a two-level system
driven around an avoided crossing, particularly in connection with
Landau-Zener (LZ) physics \cite{Landau}. These avoided crossing
regions have a special significance in QIP applications because
coherence times are usually longest in those regions, hence the
term optimal point. Needless to say, nontrivial evolution of
quantum systems is usually associated with some kind of
energy-level crossing.

In this paper we discuss the dynamics of a two-level system driven
by strong ac fields in the vicinity of an avoided crossing. As
mentioned above, numerous studies have been devoted to this
problem, approaching it from different angles and applying it to
different physical systems. Here we present a geometric picture
that leads to a simple understanding of the behaviour of the
system. This approach can also be used to derive quantitative
results in regions where other methods fail. We also keep in mind
the idea of trying to find useful applications for strong driving
as a tool to manipulate two-level systems, or as they are referred
to in the context of QIP, qubits (in this context, see
e.g.~\cite{Oliver,Sillanpaa,Shytov,Shevchenko}).

A good example displaying the richness of strongly driven
two-level systems is the so-called coherent destruction of
tunnelling (CDT) \cite{Grossmann}. A particle in a symmetric
double-well potential will generally oscillate back and forth
between the two wells. If we now add an oscillating energy
difference between the two wells, the frequency of the tunnelling
oscillations changes. At certain combinations of the driving
parameters, the tunnelling oscillations are frozen. This
phenomenon has been analyzed from several different perspectives
\cite{Kayanuma93,Garraway,Grifoni,Creffield,Floquet,Kayanuma00}.
We shall show below that it can be understood rather easily using
our approach.

Here we consider the situation where we start with an undriven
system at an arbitrary bias point. We then analyze the dynamics
that results from the application of strong driving fields. As one
would expect, resonance peaks occur at properly chosen values of
the parameters. We analyze the resonance conditions using two
approaches: one employing a rotating-wave approximation (RWA) and
one employing an approximation of discretized evolution
characterized by a sequence of fast LZ crossings. Between the two
approaches, and the well-known weak-driving case, most of the
parameter space is covered.

This paper is organized as follows. In Sec.~II we present the
model system, the Hamiltonian that describes it and some
preliminary arguments. In Sec.~III we present a RWA that can be
used to describe the dynamics in a certain region of the parameter
space. In Sec.~IV we present a geometric picture that is useful to
describe the dynamics in another region of the parameter space
(note that there is some overlap between the validity regions of
Secs. III and IV). Section V contains a discussion of the results
and some concluding remarks.

\section{Model system and Hamiltonian}

We consider a two-level system described by the Hamiltonian:
\begin{equation}
\hat{H} (t) = -\frac{\Delta}{2} \hat{\sigma}_x -
\frac{\epsilon(t)}{2} \hat{\sigma}_z.
\label{eq:Hamiltonian}
\end{equation}
where $\Delta$ is the (time-independent) coupling strength between
the two basis states \cite{Basis}; $\epsilon(t)$ is the
time-dependent bias point; and $\hat{\sigma}_x$ and
$\hat{\sigma}_z$ are the usual $x$ and $z$ Pauli matrices,
respectively. Note that we take $\hbar=1$ throughout this paper.
For definiteness and simplicity in the algebra, we assume harmonic
driving, i.e. we assume that $\epsilon(t)$ can be expressed as
\begin{equation}
\epsilon(t) = \epsilon_0 + A \cos (\omega t + \phi)
\label{eq:DrivingField}
\end{equation}
where $\epsilon_0$ is the dc component of the bias point; and $A$,
$\omega$ and $\phi$ are, respectively, the amplitude, frequency
and phase of the driving field. We shall, with no loss of
generality, take all the parameters in Eqs.~(\ref{eq:Hamiltonian})
and (\ref{eq:DrivingField}) to be positive. In order to simplify
the appearance of the expressions below, we shall take $\phi=0$.
This only simplifies the intermediate steps of the algebra, but it
does not affect any of the main results. The energy-level diagram
and the applied driving field are depicted in Fig.~1.

\begin{figure}[h]
\includegraphics[width=6.0cm]{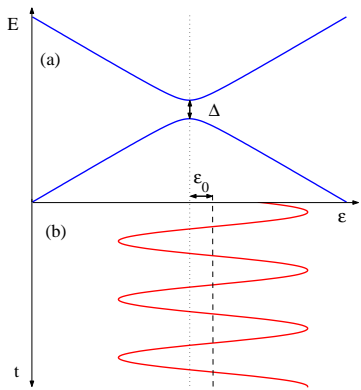}
\caption{(color online) (a) Energy-level diagram $E(\epsilon)$, in
blue, of a two level system with minimum gap $\Delta$ and (b) the
time-periodic bias $\epsilon(t)$, in red. The vertical dashed line
represents dc bias point $\epsilon_0$.}
\end{figure}

We should mention here that recently this setup was studied
experimentally using superconducting qubits in
Refs.~\cite{Oliver,Sillanpaa}, both of which contain theoretical
analysis that overlaps with ours. Related theoretical studies can
also be found in Refs.~\cite{Shytov,Shevchenko}.

In the absence of driving, the behaviour of the system is simple.
In the unbiased case (i.e., when $\epsilon_0=0$), the eigenstates
of the Hamiltonian are the symmetric and antisymmetric
superpositions of the states $\left|\uparrow\right\rangle$ and
$\left|\downarrow\right\rangle$. As a result, if the system is
initially in one of these two states, it oscillates back and forth
between them. In the strongly biased case (i.e., when
$\epsilon_0\gg\Delta$), the states $\left|\uparrow\right\rangle$
and $\left|\downarrow\right\rangle$ are, to a good approximation,
the eigenstates of the Hamiltonian. If the system is initially in
one of them, it will only experience small oscillations, occupying
the other state with a maximum probability
$(\Delta/\epsilon_0)^2$. One could say that in this case the
oscillations, which are driven by the $\hat{\sigma}_x$ term in the
Hamiltonian, are suppressed by the energy mismatch $\epsilon_0$
between the states $\left|\uparrow\right\rangle$ and
$\left|\downarrow\right\rangle$.

In the weak driving limit (i.e., when
$A\ll\sqrt{\Delta^2+\epsilon_0^2}$), Rabi-oscillation physics
applies. Resonance occurs when
$\omega=\sqrt{\Delta^2+\epsilon_0^2}$, and the frequency of Rabi
oscillations is given by $\Omega=(A \sin\alpha)/2$, where the
angle $\alpha$ is defined by the condition
$\tan\alpha=\Delta/\epsilon_0$. Rabi oscillations occur between
the eigenstates of the Hamiltonian (excluding the driving term),
such that in the unbiased case the oscillations occur between the
symmetric and antisymmetric superpositions of the states
$\left|\uparrow\right\rangle$ and $\left|\downarrow\right\rangle$.
Higher-order processes, with
$n\omega=\sqrt{\Delta^2+\epsilon_0^2}$, can be described easily as
well in this limit. We shall collectively refer to the
weak-driving limit using the easily recognizable name of
Rabi-physics limit.

Since Rabi physics is well known \cite{Baym}, we shall not discuss
it in any detail here. Instead, we focus on the case where the
amplitude $A$ is comparable to or larger than
$\sqrt{\Delta^2+\epsilon_0^2}$. A suitable RWA can be used to
obtain a good description of the dynamics when $\omega\gg\Delta$.
This approach, described in Sec.~III, can explain some interesting
features of the problem, in particular the phenomenon of CDT
mentioned in Sec.~I.

A different approximation can be used when
$(A-\epsilon_0)\gg\Delta$ and $A\omega\gg \Delta^2$ (note that
this condition overlaps with the validity condition of the RWA).
In this case we can think of the dynamics as being composed of a
sequence of LZ crossings separated by periods of free evolution of
the basis states. We shall take this approach to analyze the
problem in Sec IV. We shall then compare the validity conditions
of the different approximations in Sec.~V.

\section{High-frequency driving: Rotating-wave approximation}

We now take the system described by the Hamiltonian in
Eq.~(\ref{eq:Hamiltonian}) and make a transformation to a rotating
frame, such that a wave function $|\psi\rangle$ in the lab frame
can be expressed as \cite{Pegg}:
\begin{equation}
|\psi\rangle = \hat{U}(t) |\psi'\rangle,
\end{equation}
where
\begin{equation}
\hat{U}(t) = \exp \left\{\frac{i}{2} \left( \epsilon_0 t +
\frac{A}{\omega} \sin \omega t \right) \hat{\sigma}_z \right\},
\end{equation}
and $|\psi'\rangle$ is the wave function in the rotating frame. In
other words, we use the interaction picture with the coupling term
treated as a perturbation. The Schr\"odinger equation
\begin{equation}
i \frac{d}{dt} |\psi\rangle = \hat{H}(t) |\psi\rangle
\end{equation}
can now be written as
\begin{equation}
i \frac{d}{dt} |\psi'\rangle = \hat{H}'(t) |\psi'\rangle,
\end{equation}
with
\begin{eqnarray}
\hat{H}'(t) & = & \hat{U}^{\dagger}(t) \hat{H}(t) \hat{U}(t) - i
\hat{U}^{\dagger}(t) \frac{ d \hat{U}(t)}{dt}
\nonumber
\\
& = & -\frac{\Delta}{2} e^{- \frac{i}{2} \left( \epsilon_0 t +
\frac{A}{\omega} \sin \omega t \right) \hat{\sigma}_z}
\hat{\sigma}_x e^{\frac{i}{2} \left( \epsilon_0 t +
\frac{A}{\omega} \sin \omega t \right) \hat{\sigma}_z}
\nonumber
\\
& = & - \frac{\Delta}{2} \left(
\begin{array}{cc}
0 & e^{ - i \left( \epsilon_0 t + \frac{A}{\omega} \sin \omega
t \right)} \\
e^{ i \left( \epsilon_0 t + \frac{A}{\omega} \sin \omega t
\right)} & 0
\end{array}
\right).
\end{eqnarray}
We now make use of the relation
\begin{equation}
\exp \left\{i z \sin \gamma \right\} = \sum_{n=-\infty}^{\infty}
J_n(z) \; e^{i n \gamma},
\end{equation}
where $J_n(x)$ are Bessel functions of the first kind, and we find
that
\begin{widetext}
\begin{eqnarray}
\hat{H}'(t) & = & - \frac{\Delta}{2} \left(
\begin{array}{cc}
0 & \displaystyle\sum_{n=-\infty}^{\infty}
J_n\left(\frac{A}{\omega}\right) e^{-i (n \omega + \epsilon_0) t}
\\
\displaystyle\sum_{n=-\infty}^{\infty}
J_n\left(\frac{A}{\omega}\right) e^{i (n \omega + \epsilon_0) t} &
0
\end{array}
\right).
\label{eq:Hprime}
\end{eqnarray}
\end{widetext}
Note that in going from Eq.~(\ref{eq:Hamiltonian}) to
Eq.~(\ref{eq:Hprime}) we have not made any approximations. It will
be useful below to use the following asymptotic behaviour of the
Bessel functions:
\begin{eqnarray}
J_n(z) & \approx & \frac{z^n}{2^n n!} , \hspace{3.75cm} z \ll 1
\nonumber
\\
J_n(z) & \approx & \sqrt{\frac{2}{\pi z}} \cos\left[
z-(2n+1)\frac{\pi}{4} \right] , \hspace{0.5cm} z \gg n.
\label{eq:asymptotics}
\end{eqnarray}

We now perform a rotating wave approximation (RWA): we assume that
all the terms in the sum in Eq.~(\ref{eq:Hprime}) oscillate very
fast compared to the timescale of the (coarse-grained, or
smoothened) system dynamics, with the exception of one term that
is identified as the resonant or near-resonant term. The effect of
the non-resonant terms can therefore be neglected when studying
the long-time system dynamics. We should note here that this RWA
is different from the one widely known and used in the case of
weak driving. In that case the linear driving field is replaced by
a rotating field, clockwise or counter-clockwise depending on
conventions. The other component of the field, i.e. the one
rotating in the opposite sense, is neglected \cite{Gardiner}.

The resonance condition is identified using the intuitive
requirement that the resonant term of the RWA is constant in time.
We therefore have the resonance condition:
\begin{equation}
n \omega + \epsilon_0 = 0
\label{eq:simpleRC}
\end{equation}
for some integer $n$. With parameters satisfying the resonance
condition, the $\left|\uparrow\right\rangle \leftrightarrow
\left|\downarrow\right\rangle$ oscillation frequency is given by:
\begin{equation}
\Omega = \Delta \left| J_n\left(\frac{A}{\omega}\right) \right|.
\label{eq:RWAOscillationFrequency}
\end{equation}
One usually identifies the resonance with a given value of $n$ as
describing an $|n|$-photon process. The Rabi resonance condition
corresponds to the case $n=-1$, where we find that
$\omega=\epsilon_0$ (the difference from the condition
$\omega=\sqrt{\Delta^2+\epsilon_0^2}$ will become clear shortly).
Assuming weak driving, we find that the frequency of Rabi
oscillations is given by:
\begin{equation}
\Omega = \frac{\Delta}{\epsilon_0} \times \frac{A}{2}.
\end{equation}
For large values of $\epsilon_0$, the factor $\Delta/\epsilon_0$
is a good approximation to the factor $\sin\alpha$, where $\alpha$
is the angle between the static bias field and the driving field.
The Rabi frequency vanishes asymptotically as $\epsilon_0
\rightarrow \infty$, as it should. Note also that $\Omega$ (Eq.
\ref{eq:RWAOscillationFrequency}) increases with increasing $A$
for small values of $A$, but generally decreases as $1/\sqrt{A}$
for large values of $A$. The mechanism responsible for this latter
behaviour will become clear in Sec.~IV. In particular, note that
the turning point between the two behaviours (or, alternatively,
the maximum in $\Omega$) occurs when $A$ is comparable to
$\epsilon_0$.

The width of the resonance can be obtained from the following
considerations. If the driving frequency is shifted from exact
resonance (Eq.~\ref{eq:simpleRC}) by $\delta\omega$, the resonant
(i.e., slowest) term in Eq.~(\ref{eq:Hprime}) oscillates with
frequency $n\delta\omega$. When these oscillations become faster
than the $\left|\uparrow\right\rangle \leftrightarrow
\left|\downarrow\right\rangle$ oscillation dynamics, which is
characterized by $\Omega$, the resonance is clearly lost. The
width of the resonance can therefore be taken as:
\begin{equation}
\delta\omega \sim \frac{\Omega}{|n|}.
\end{equation}
Using higher-order processes (i.e., with $|n|>1$) therefore
results in resonances that are narrow compared to the on-resonance
oscillation frequency. This property can be useful, for example,
in applications where one is dealing with several closely spaced
resonances. If one is trying to drive only one of those
resonances, this approach provides a possibility to target a
single resonance, without necessarily making the oscillation
dynamics extremely slow.

We now note that for the above resonance condition
(Eq.~\ref{eq:simpleRC}) to hold we must be able to neglect all the
oscillating terms. This means that we require that
$|n\omega+\epsilon_0|\gg \Delta$ for all $n$ except the one
satisfying the resonance condition \cite{DroppingBessel}. Keeping
in mind that $|n\omega+\epsilon_0|=0$ for one value of $n$, we
find that the validity condition can be expressed simply as
$\omega\gg\Delta$. This explains the difference between the above
resonance condition and the usual resonance condition for Rabi
oscillations; for example, when $\epsilon_0=0$ and $\omega=\Delta$
we cannot keep only one term in Eq.~(\ref{eq:Hprime}), and this
approximation breaks down. Note, however, that even if
$\epsilon_0=0$ we can still use this approximation, as long as
$\omega\gg\Delta$. This approach is very useful when discussing
the phenomenon of CDT, which we do next.

The much-analyzed phenomenon of CDT \cite{Grossmann} can be
explained using the above approach. In the unbiased case (i.e.,
when $\epsilon_0=0$), Eq.~(\ref{eq:simpleRC}) gives the
interesting result that, regardless of the value of $\omega$,
taking $n=0$ always satisfies the resonance condition. This means
that oscillations between the states $\left|\uparrow\right\rangle$
and $\left|\downarrow\right\rangle$ will always occur with full
$\left|\uparrow\right\rangle \leftrightarrow
\left|\downarrow\right\rangle$ conversion, provided of course that
we can neglect all the terms with $n\neq 0$ (or in other words
$\omega\gg\Delta$). This statement is obvious for no driving
($A=0$), but it is not obvious that for large driving fields
($A\gg \Delta$) full conversion should occur. From this point of
view, it looks more surprising that full oscillations occur at all
for strong driving, even though the system hardly spends any time
in the degeneracy region \cite{AsymptoticFrequency}. Accepting the
existence of these oscillations, we now take the oscillation
frequency as given by Eq.~(\ref{eq:RWAOscillationFrequency}) with
$n=0$. It is now clear that CDT occurs when
\begin{equation}
J_0\left(\frac{A}{\omega}\right)=0,
\label{eq:CDTcondition}
\end{equation}
such that the resonant term in Eq.~(\ref{eq:Hprime}) has a
vanishing coefficient (note that Eq.~(\ref{eq:CDTcondition})
agrees with the results of previous work
\cite{Grossmann,Kayanuma93}). With parameters satisfying the CDT
condition, the frequency of $\left|\uparrow\right\rangle
\leftrightarrow \left|\downarrow\right\rangle$ oscillations
vanishes, and the oscillations are consequently suppressed. One
could therefore say that CDT is simply the statement that the
$\left|\uparrow\right\rangle \leftrightarrow
\left|\downarrow\right\rangle$ oscillation frequency becomes
extremely small if we use driving parameters that are close to a
point satisfying Eq.~(\ref{eq:CDTcondition}).

\section{Repeated traversals of the crossing region: Transfer-matrix (TM) approach}

We now focus on the case of strong driving where the system
repeatedly traverses the crossing region [i.e.,
$(A-\epsilon_0)\gg\Delta$]. Even in the case $\omega\gg\Delta$,
where the treatment using the RWA above is still valid, that
approach becomes less intuitive. Instead, one can gain better
insight into the problem by analyzing the dynamics as composed of
discrete steps, as we shall do in this section (see also
Ref.~\cite{Sillanpaa}).

Let us take the limit where $(A-\epsilon_0)\gg\Delta$. We can now
think of the system as undergoing a sequence of LZ crossings
(represented by the red and green lines in Fig.~2). At each
crossing event, the states $\left|\uparrow\right\rangle$ and
$\left|\downarrow\right\rangle$ experience some mixing. If we
approximate the sweep across the degeneracy region by a linear
ramp of the bias point between two points located symmetrically
around the degeneracy point (denoted by $t_c-\tau$ and $t_c+\tau$,
where $t_c$ is the time at which the bias point is at the center
of the avoided crossing), we find that the LZ crossing can be
approximately described by the evolution matrix:
\begin{equation}
\hat{G}_{\rm LZ, \it k} = \left( \begin{array}{cc}
\cos \displaystyle\frac{\chi}{2} & \sin \displaystyle\frac{\chi}{2} e^{i\theta_{\rm LZ, \it k}} \\
-\sin \displaystyle\frac{\chi}{2} e^{-i\theta_{\rm LZ, \it k}} &
\cos \displaystyle\frac{\chi}{2}
\end{array} \right),
\label{eq:TransferMatrix}
\end{equation}
where the angle $\chi$ is defined by the LZ transition probability
\begin{eqnarray}
\sin^2 \displaystyle\frac{\chi}{2} & = & 1 - \exp \left\{
-\frac{\pi\Delta^2}{2v} \right\},
\end{eqnarray}
$v$ is the sweep rate, and $k$ defines the direction of the bias
sweep across the degeneracy region: $k=1$ when $\epsilon(t)$ goes
from positive values to negative values and $k=2$ when
$\epsilon(t)$ goes from negative values to positive values (Note
that the sweep rates in the two directions are equal, and thus
$\chi$ is independent of $k$). The angles $\theta_{\rm LZ, \it k}$
are given by
\begin{eqnarray}
\theta_{\rm LZ, 1} & = & \pi - \theta_{\rm Stokes} - \frac{1}{2}
\int_{0}^{\tau} \sqrt{\Delta^2+(A^2-\epsilon_0^2)\omega^2\tau^2}
d\tau
\nonumber
\\
& = & \pi - \theta_{\rm Stokes} - \frac{1}{2} \tau
\sqrt{\Delta^2+(A^2-\epsilon_0^2)\omega^2\tau^2} -
\nonumber
\\
& & \frac{\Delta^2}{2\sqrt{A^2-\epsilon_0^2} \omega} \sinh^{-1}
\frac{\sqrt{A^2-\epsilon_0^2} \omega \tau}{\Delta}
\nonumber
\\
& \approx & \pi - \theta_{\rm Stokes} - \frac{1}{2}
\sqrt{A^2-\epsilon_0^2} \omega \tau^2 -
\nonumber
\\
& & \frac{\Delta^2}{2\sqrt{A^2-\epsilon_0^2} \omega} \log
\frac{2\sqrt{A^2-\epsilon_0^2} \omega \tau}{\Delta}
\nonumber
\\
\theta_{\rm LZ, 2} & \approx & \theta_{\rm Stokes} + \frac{1}{2}
\sqrt{A^2-\epsilon_0^2} \omega \tau^2 +
\nonumber
\\
& & \frac{\Delta^2}{2\sqrt{A^2-\epsilon_0^2} \omega} \log
\frac{2\sqrt{A^2-\epsilon_0^2} \omega \tau}{\Delta}
\nonumber
\\
\theta_{\rm Stokes} & = & \frac{\pi}{4} + {\rm arg} \left[ \Gamma
(1-i\delta) \right] + \delta ( \ln \delta - 1 )
\nonumber
\\
\delta & = & \frac{\Delta^2}{4v},
\label{eq:LZphases}
\end{eqnarray}
and $\Gamma(x)$ is the gamma function (see
e.g.~Ref.~\cite{ShevchenkoReview} for a detailed derivation). Note
that the Stokes phase $\theta_{\rm Stokes}$ approaches zero in the
slow-crossing limit (i.e., when $\delta\rightarrow \infty$), and
it approaches $\pi/4$ in the fast-crossing limit (i.e., when
$\delta\rightarrow 0$) \cite{Berry}. The description of LZ
crossing processes using unitary matrices of the form of
Eq.~(\ref{eq:TransferMatrix}) is sometimes referred to as the
transfer-matrix method, which we follow here (the term scattering
matrix can also be found in the literature).

Between each two LZ crossings, the system moves far from the
degeneracy region and acquires a relative-phase factor between the
states $\left|\uparrow\right\rangle$ and
$\left|\downarrow\right\rangle$, but with no mixing between them.
Because of the asymmetry (i.e., the fact that, in general,
$\epsilon_0\neq 0$), there are two phase factors corresponding to
the system being on the right or left side of the degeneracy
region. We therefore find that between crossings, the system
evolves by the evolution matrices:
\begin{equation}
\hat{G}_{j} = \left( \begin{array}{cc}
e^{-i\theta_j} & 0 \\
0 & e^{i\theta_j}
\end{array} \right),
\end{equation}
where $j$ represents the two sides of the degeneracy point (we
shall refer to them as 1 and 2).

Before starting to analyze the dynamics resulting from combining
the above matrices, we note that the values of the angles
$\theta_1$, $\theta_2$, $\theta_{\rm LZ,1}$ and $\theta_{\rm
LZ,2}$ depend on where we set the boundaries between the different
time steps (i.e.~they depend on $\tau$). It is therefore useful to
define the boundary-independent phase factors $\tilde{\theta}_1$,
$\tilde{\theta}_2$, $\tilde{\theta}_{\rm LZ,1}$ and
$\tilde{\theta}_{\rm LZ,1}$. We define $\tilde{\theta}_1$ and
$\tilde{\theta}_2$ as half the relative phases accumulated between
the energy eigenstates between the times of two successive level
crossings (ignoring in this evaluation the fact that there is
mixing dynamics occurring between the states):
\begin{eqnarray}
\tilde{\theta}_1 & = & - \int \frac{1}{2}
\sqrt{(\epsilon_0+A\cos\omega t)^2 + \Delta^2} dt
\nonumber
\\
& = & - \int \frac{1}{2} (\epsilon_0+A\cos\omega t) dt -
f_1(\epsilon_0,A,\omega,\Delta)
\nonumber
\\
& = & - \frac{\sqrt{A^2-\epsilon_0^2}}{\omega} -
\frac{\epsilon_0}{\omega} \cos^{-1}\frac{-\epsilon_0}{A} -
f_1(\epsilon_0,A,\omega,\Delta)
\nonumber
\\
& = & - \frac{\sqrt{A^2-\epsilon_0^2}}{\omega} +
\frac{\epsilon_0}{\omega} \cos^{-1}\frac{\epsilon_0}{A} -
\frac{\pi \epsilon_0}{\omega} - f_1(\epsilon_0,A,\omega,\Delta),
\nonumber
\\
\tilde{\theta}_2 & = & \frac{\sqrt{A^2-\epsilon_0^2}}{\omega} -
\frac{\epsilon_0}{\omega} \cos^{-1}\frac{\epsilon_0}{A} +
f_2(\epsilon_0,A,\omega,\Delta),
\label{eq:theta_tildes}
\end{eqnarray}
where $f_1$ and $f_2$ are functions that, roughly speaking, are
proportional to $\Delta^2/(A\omega)$ times the logarithm of
$A/\Delta$ (see Eq.~\ref{eq:LZphases}). With this definition of
$\tilde{\theta}_1$ and $\tilde{\theta}_2$, we find that the
appropriate values of $\tilde{\theta}_{\rm LZ,1}$ and
$\tilde{\theta}_{\rm LZ,2}$ are given by:
\begin{eqnarray}
\tilde{\theta}_{\rm LZ,1} & = & \pi - \theta_{\rm Stokes}
\nonumber
\\
\tilde{\theta}_{\rm LZ,2} & = & \theta_{\rm Stokes}.
\end{eqnarray}

\begin{figure}[h]
\includegraphics[width=6.0cm]{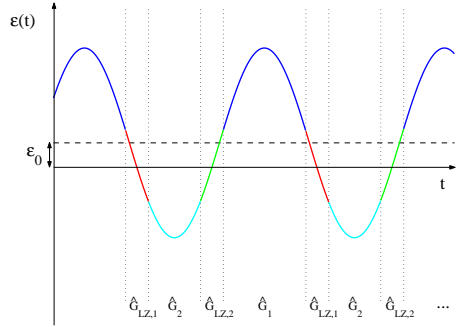}
\caption{(color online) The time evolution of the system divided
into discrete time steps, each of which can be described by a
simple evolution matrix.}
\end{figure}

We now have a sequence of finite time steps, each of which is
described by a simple evolution matrix, as shown in Fig.~2. The
dynamics over a large number of driving cycles can be constructed
by multiplying the evolution matrices. The evolution of the system
over $N$ cycles is therefore described by the matrix
\begin{equation}
\hat{G}_N = \left( \hat{G}_{\rm LZ,2} \hat{G}_{2} \hat{G}_{\rm
LZ,1} \hat{G}_{1} \right)^N,
\label{eq:G_N}
\end{equation}
with possibly some minor modifications at the beginning and end of
the sequence, depending on the exact initial and final bias points
\cite{SlowDynamics}. We now take a single cycle and calculate the
evolution matrix that describes it. We find that
\begin{equation}
\hat{G}_{\rm LZ,2} \; \hat{G}_{2} \; \hat{G}_{\rm LZ,1} \;
\hat{G}_{1} = \left(
\begin{array}{cc}
g_{11} & g_{12} \\
-g_{12}^* & g_{11}^*
\end{array} \right),
\label{eq:DiscretePropagator}
\end{equation}
where
\begin{widetext}
\begin{eqnarray}
g_{11} & = & \cos^2 \displaystyle\frac{\chi}{2} \; e^{-i(\theta_1
+ \theta_2)} - \sin^2 \displaystyle\frac{\chi}{2} \;
e^{i(-\theta_{\rm LZ,1}+\theta_{\rm LZ,2}-\theta_1+\theta_2)}
\nonumber
\\
g_{12} & = & \sin \displaystyle\frac{\chi}{2} \cos
\displaystyle\frac{\chi}{2} \left( e^{i(\theta_{\rm LZ,1} +
\theta_1 - \theta_2)} + e^{i(\theta_{\rm LZ,2} + \theta_1 +
\theta_2)} \right)
\end{eqnarray}
It is useful to rewrite Eq.~(\ref{eq:DiscretePropagator}) as
\begin{equation}
\hat{G}_{\rm LZ,2} \; \hat{G}_{2} \; \hat{G}_{\rm LZ,1} \;
\hat{G}_{1} = \left(
\begin{array}{cc}
\cos\displaystyle\frac{\zeta_{\rm FC}}{2} &
\sin\displaystyle\frac{\zeta_{\rm FC}}{2} \; e^{i\phi_{\rm FC}} \\
-\sin\displaystyle\frac{\zeta_{\rm FC}}{2} \; e^{-i\phi_{\rm FC}}
& \cos\displaystyle\frac{\zeta_{\rm FC}}{2}
\end{array} \right) \hspace{0.1cm}
\left(
\begin{array}{cc}
e^{-i \theta_{\rm FC}/2} & 0 \\
0 & e^{i \theta_{\rm FC}/2}
\end{array} \right),
\label{eq:DiscretePropagatorSeparated}
\end{equation}
where the subscript FC indicates that the above equation describes
the evolution of the system over a full cycle of the driving
field. We now find that
\begin{eqnarray}
\sin^2\displaystyle\frac{\zeta_{\rm FC}}{2} & = & 4 \sin^2
\displaystyle\frac{\chi}{2} \cos^2
\left(\frac{\theta_{LZ,1}-\theta_{LZ,2}}{2}-\theta_2 \right)
\nonumber
\\
\theta_{\rm FC} & = & 2 \arctan \frac{\cos^2 \chi/2 \sin
(\theta_1+\theta_2) + \sin^2 \chi/2 \sin
(\theta_{LZ,1}-\theta_{LZ,2}+\theta_1-\theta_2)}{\cos^2 \chi/2
\cos (\theta_1+\theta_2) - \sin^2 \chi/2 \cos
(\theta_{LZ,1}-\theta_{LZ,2}+\theta_1-\theta_2)}
\nonumber
\\
\phi_{\rm FC} & = & \frac{\theta_{LZ,1}+\theta_{LZ,2} + 2 \theta_1
- \theta_{\rm FC}}{2}.
\end{eqnarray}
\end{widetext}

We can now analyze the dynamics described by
Eq.~(\ref{eq:DiscretePropagatorSeparated}). The first matrix in
the product describes a rotation by a some angle $\zeta_{\rm FC}$
around an axis in the $xy$-plane. The second matrix in the product
describes a rotation by an angle $\theta_{\rm FC}$ around the $z$
axis. Since we are looking for resonance-like dynamics (which
naturally implies the existence of a resonance condition), we want
the angle $\zeta_{\rm FC}$ to be small. This condition is
satisfied in two opposite limits: the fast- and slow-crossing
limits. We shall treat these two limits separately below, after
making the following general observations.

With the above geometrical interpretation of the roles of the
angles $\zeta_{\rm FC}$ and $\theta_{\rm FC}$, the resonance
condition is clear. If $\theta_{\rm FC}$ is a multiple of $2\pi$,
the $z$-axis rotation does not affect the dynamics, and the small
rotations of $\zeta_{\rm FC}$ add up to produce full oscillations
between the states $\left|\uparrow\right\rangle$ and
$\left|\downarrow\right\rangle$. If, on the other hand,
$\theta_{\rm FC}$ takes a value that is different from any
multiple of $2\pi$ by more than $\zeta_{\rm FC}$, the small
rotations of $\zeta_{\rm FC}$ will not add up in an ideal manner,
and the oscillations will be suppressed \cite{GeometricArgument}.
In the following subsections we treat the two limits where simple
analytic formulae can be obtained.

The width of the resonance can also be obtained using the
geometrical picture explained above. We require that the phase
factor $\theta_{FC}$ be within a distance $\zeta_{\rm FC}$ from a
multiple of $2\pi$. Starting from Eq.~(\ref{eq:thetaFC}), and
writing
\begin{equation}
\theta_{\rm FC} \approx 2\pi n - \frac{2\pi \epsilon_0}{\omega^2}
\; \delta \omega,
\end{equation}
where $\delta \omega$ is the deviation from exact resonance, we
find that the width of the resonance is given by:
\begin{eqnarray}
\delta\omega & \sim & \frac{\omega^2 \zeta_{\rm FC}}{2 \pi
\epsilon_0}
\nonumber
\\
& \sim & \frac{\Omega}{n}
\end{eqnarray}
As in Sec.~III, we find that in addition to the usual factor of
oscillation frequency, the width of the resonance now contains the
factor $1/n$. This means that with the proper choice of
parameters, the width of the resonance can be made substantially
smaller than the on-resonance oscillation frequency.

\subsection{Fast-crossing limit}

In this subsection we assume that each crossing is traversed in
the fast limit. This means that we require the sweep rate across
the degeneracy region (i.e., $\omega\sqrt{A^2-\epsilon_0^2}$) to
be much larger than the square of the gap size of the crossing
$\Delta^2$. The LZ probability in this limit is given by
\begin{equation}
\sin^2 \displaystyle\frac{\chi}{2} \approx \frac{\pi\Delta^2}{2v}.
\end{equation}
Note that in  this limit $\sin^2(\chi/2) \ll \cos^2(\chi/2)$.

The resonance condition for the constructive accumulation of small
rotations can now be obtained by noting that:
\begin{eqnarray}
\theta_{\rm FC} & \approx & 2 (\theta_1 + \theta_2 )
\nonumber
\\
& \approx & -\int_{\tau}^{\tau+2\pi/\omega} dt \left[ \epsilon_0 +
A \cos \omega t \right]
\nonumber
\\
& = & - \frac{2 \pi \epsilon_0}{\omega}.
\label{eq:thetaFC}
\end{eqnarray}
The resonance condition is therefore given by
\begin{equation}
\frac{\epsilon_0}{\omega} = n
\label{eq:LZresonance_condition}
\end{equation}
for some integer $n$. This is the same condition that we found in
Sec.~III, using an entirely different approach (note also that the
validity conditions of the two approaches are different, a point
to which we shall come back in Sec.~V). The angle $\phi_{\rm FC}$
is given by
\begin{equation}
\phi_{\rm FC} \approx \frac{\pi}{2} - \tilde{\theta}_2.
\end{equation}

Using the approach of this section it might seem somewhat
surprising that the amplitude $A$ does not appear in the resonance
condition, even though $A$ can be the largest energy scale in the
problem. One should also note here that Eq.~(\ref{eq:simpleRC}) is
only approximate in the weak-driving limit, with a correction that
is proportional to $A^2$ (i.e. the Bloch-Siegert shift). One might
therefore expect that such corrections will take over at some
point, such that the amplitude $A$ becomes an essential part of
the resonance condition. Counterintuitively, however, these
corrections change behaviour and vanish asymptotically for large
$A$ (in the fast-crossing limit). The fact that the two
(identical) resonance conditions are good approximations deep in
opposite limits demonstrates further that the derivation of
Eq.~(\ref{eq:LZresonance_condition}) in this section should not be
thought of as simply a re-derivation of Eq.~(\ref{eq:simpleRC}).
It is worth noting here that even for non-harmonic driving we can
follow a similar analysis to what was done in this section and
find that the resonance condition is still independent of the
driving amplitude, with $\epsilon_0$ replaced by the time-averaged
value of the bias point and $\omega$ replaced by $2\pi$ over the
driving period [see Eq.~(\ref{eq:thetaFC})].

The frequency of oscillations on resonance can now be obtained
rather straightforwardly. If we assume that the resonance
condition in Eq.~(\ref{eq:LZresonance_condition}) is satisfied, we
find that the $\left|\uparrow\right\rangle\leftrightarrow
\left|\downarrow\right\rangle$ oscillation frequency is given by:
\begin{eqnarray}
\Omega & = & \frac{\omega \zeta_{FC}}{2\pi}
\nonumber
\\
& \approx & \frac{2 \omega}{\pi} \sin\displaystyle\frac{\chi}{2}
\left| \cos \left( \frac{\theta_{\rm LZ,1}-\theta_{\rm LZ,2}}{2} -
\theta_2 \right) \right|.
\label{eq:TMOscillationFrequencyDerivation}
\end{eqnarray}
We therefore find that
\begin{equation}
\Omega \approx \frac{2\omega}{\pi} \sqrt{\frac{\pi \Delta^2}{2
\omega \sqrt{A^2-\epsilon_0^2}}} \left| \cos \left(
\tilde{\theta}_2 - \frac{\pi}{4} \right) \right|.
\label{eq:TMOscillationFrequency}
\end{equation}
In the fast-crossing limit, the functions $f_1$ and $f_2$ in
Eq.~(\ref{eq:theta_tildes}) are negligibly small, and the phase
$\tilde{\theta}_2$ is approximately given by
\begin{equation}
\tilde{\theta}_2 \approx \frac{\sqrt{A^2-\epsilon_0^2}}{\omega} -
\frac{\epsilon_0}{\omega} \cos^{-1}\frac{\epsilon_0}{A}.
\label{eq:theta2tilde}
\end{equation}
In the special case $\epsilon_0=0$, we find that
$\tilde{\theta}_2=A/\omega$, and we recover
Eq.~(\ref{eq:RWAOscillationFrequency}) with $n=0$ for the
oscillation frequency.

One might wonder why several expressions above are not symmetric
with respect to $\theta_1$ and $\theta_2$. This asymmetry results
from our grouping of evolution matrices into full cycles [see
Eq.~(\ref{eq:G_N})], as well as the order of matrices in
Eq.~(\ref{eq:DiscretePropagatorSeparated}). These are clearly
matters of convention. Different orderings of the matrices can
result in different-looking expressions. The end result must of
course be independent of this choice. For example, if we
substitute Eq.~(\ref{eq:theta2tilde}) into
Eq.~(\ref{eq:TMOscillationFrequency}), we do not see any
convention dependence.

In order to illustrate the transfer-matrix picture, we show in
Figs. 3-5 numerical simulations of the dynamics in the validity
region of that picture. We plot the occupation probability of the
state $\left|\uparrow\right\rangle$ as a function of time,
assuming that the system was initially in the state
$\left|\downarrow\right\rangle$. We can see in Fig.~3 that the
occupation probability exhibits sudden jumps that correspond to LZ
crossings. The steps are rather large in this figure because the
crossings are not in the fast limit. If we look on long time
scales, we can see that the dynamics looks like sinusoidal
oscillations. This long-time behaviour becomes particularly smooth
when the transition probability in a single LZ crossing is small,
as is the case in Fig.~4. We also plot in Figs.~4 and 5
sinusoidally oscillating functions with frequencies given by
Eq.~(\ref{eq:TMOscillationFrequency}) \cite{PlottedTM}, as well as
sinusoidally oscillating functions with frequencies given by
Eq.~(\ref{eq:RWAOscillationFrequency}) from Sec.~III. In Fig.~5
the driving frequency $\omega$ is smaller than $\Delta$. We
therefore find inconsistency in the predictions of
Eq.~(\ref{eq:RWAOscillationFrequency}). As a general rule, the RWA
gives reasonable or good agreement with the numerical simulations
when the resulting oscillation frequency in the system dynamics is
large. When the oscillation frequency is small, the effect of the
resonant term is not necessarily large compared to that of the
other terms in Eq.~(\ref{eq:Hprime}), and the RWA fails. This is
most clearly seen in Fig.~5(c). In Fig.~4, we are in the region
where $\omega>\Delta$, and we always find that both
Eq.~(\ref{eq:RWAOscillationFrequency}) and
Eq.~(\ref{eq:TMOscillationFrequency}) agree well with the
numerical simulations. It should also be noted that since Fig.~5
corresponds to parameters that are not deep in the fast-crossing
limit, Eq.~(\ref{eq:TMOscillationFrequency}) shows some deviation
from the true oscillation frequency (see Fig.~5a).

\begin{figure}[h]
\includegraphics[width=6.5cm]{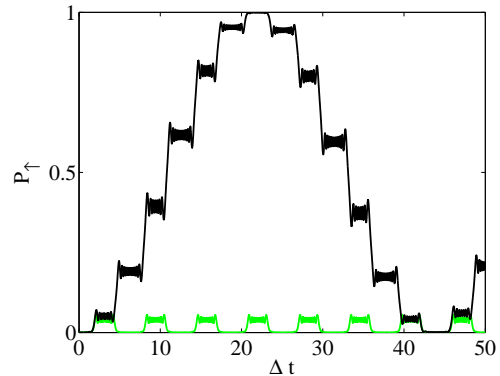}
\caption{(color online) Occupation probability $P_{\uparrow}$ of
the state $\left|\uparrow\right\rangle$ as a function of time (in
dimensionless units), assuming that the system was initially in
the state $\left|\downarrow\right\rangle$. We take
$\epsilon_0/\Delta=5$, and $\omega/\Delta=1$. The driving
amplitude is given by $A/\Delta=30$ for the black line and
$A/\Delta=34.95$ for the green (gray) line [note that the
resonance condition is satisfied in both cases]. The
transfer-matrix method and the RWA of Sec.~III both agree well
with the numerical results (their predictions are not shown here
for clarity).}
\end{figure}
\begin{figure}[h]
\includegraphics[width=6.5cm]{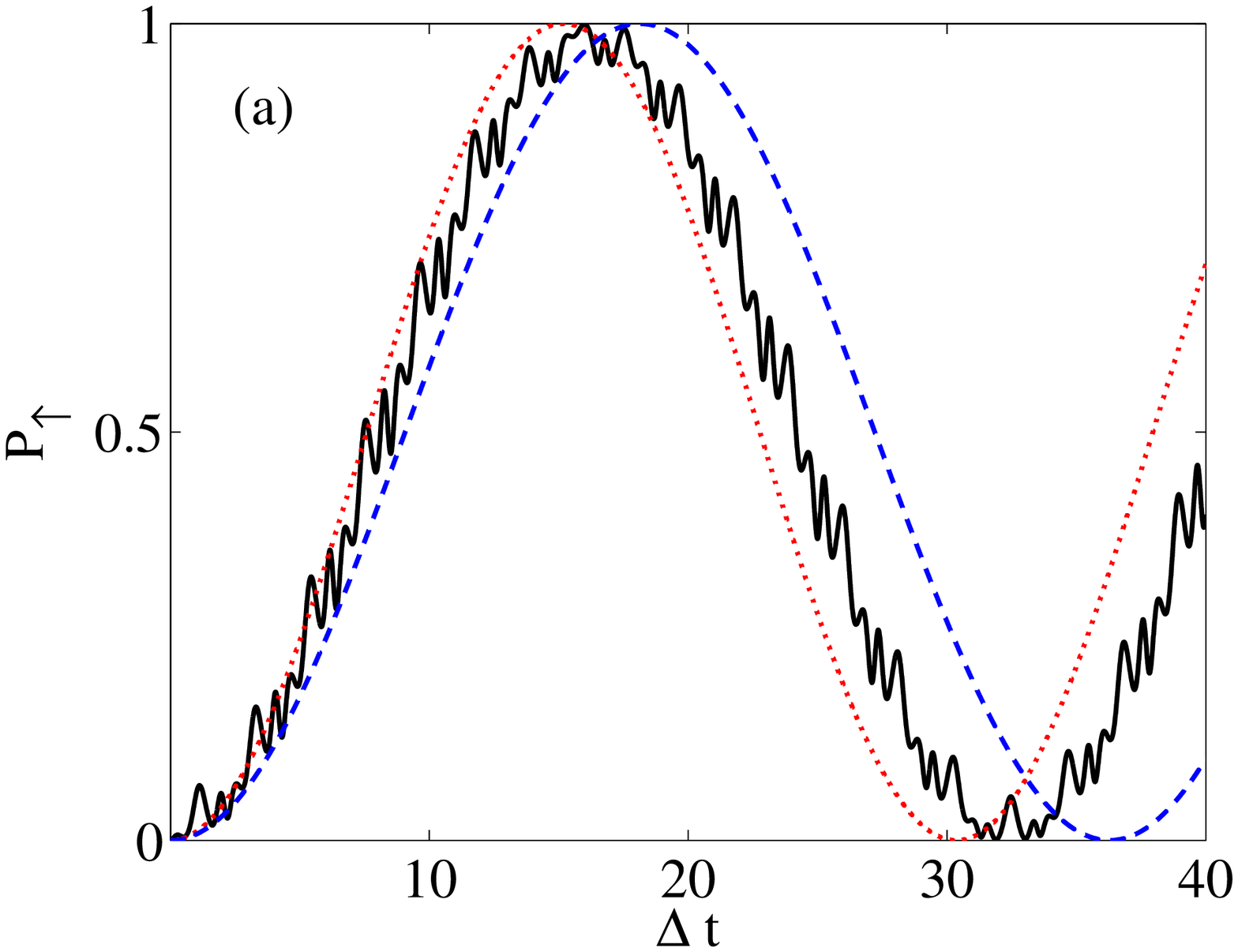}
\includegraphics[width=6.5cm]{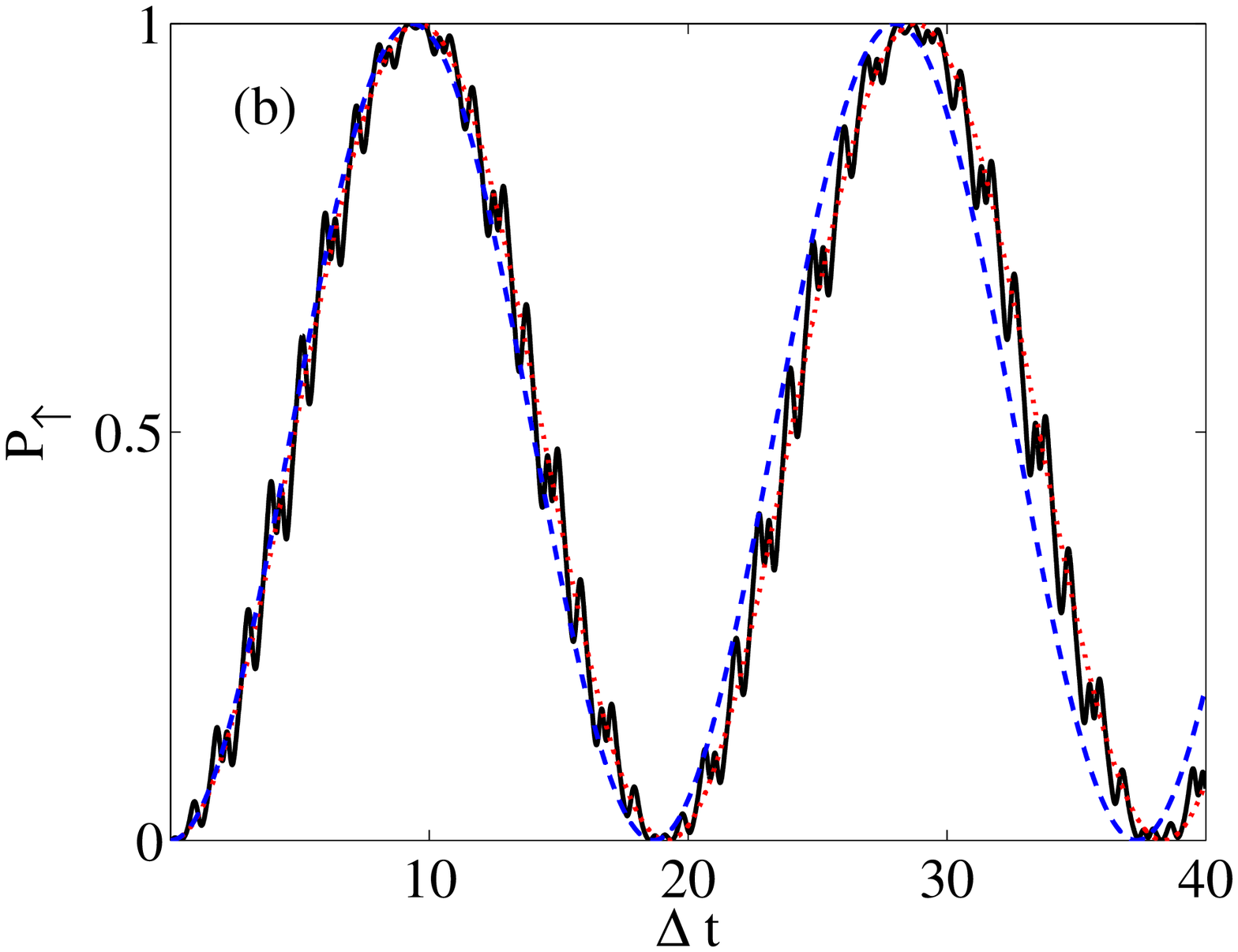}
\includegraphics[width=6.5cm]{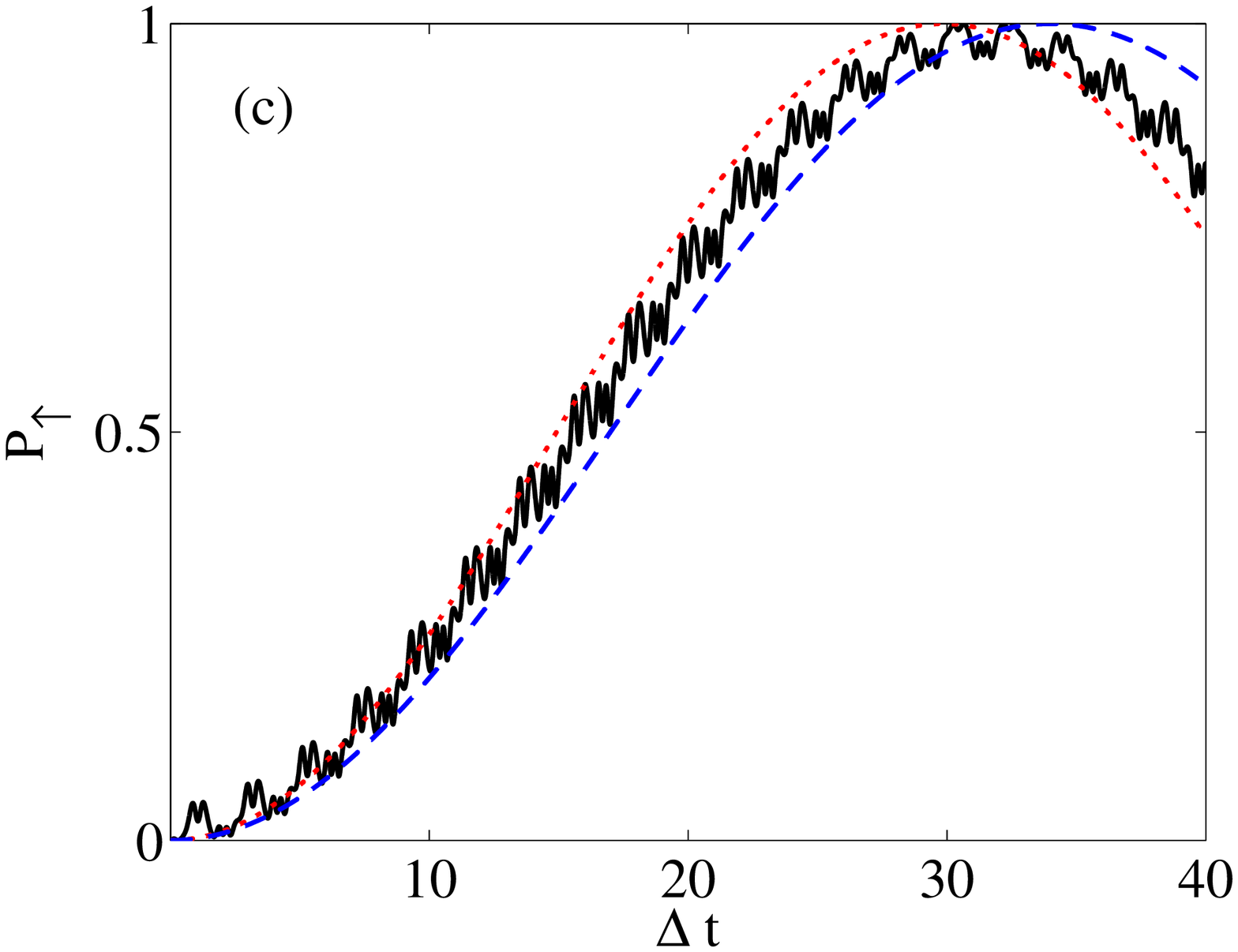}
\caption{(color online) Occupation probability $P_{\uparrow}$ of
the state $\left|\uparrow\right\rangle$ as a function of time (in
dimensionless units), assuming that the system was initially in
the state $\left|\downarrow\right\rangle$. In all the figures
$\epsilon_0/\Delta=\omega/\Delta=3$. The driving amplitude is
given by $A/\Delta=$10(a), 15(b) and 20(c). The blue dashed line
gives the (coarse-grained) predicted dynamics from the
transfer-matrix method in the large-amplitude limit, and the red
dotted line gives the predicted dynamics from Sec.~III.}
\end{figure}
\begin{figure}[h]
\includegraphics[width=6.5cm]{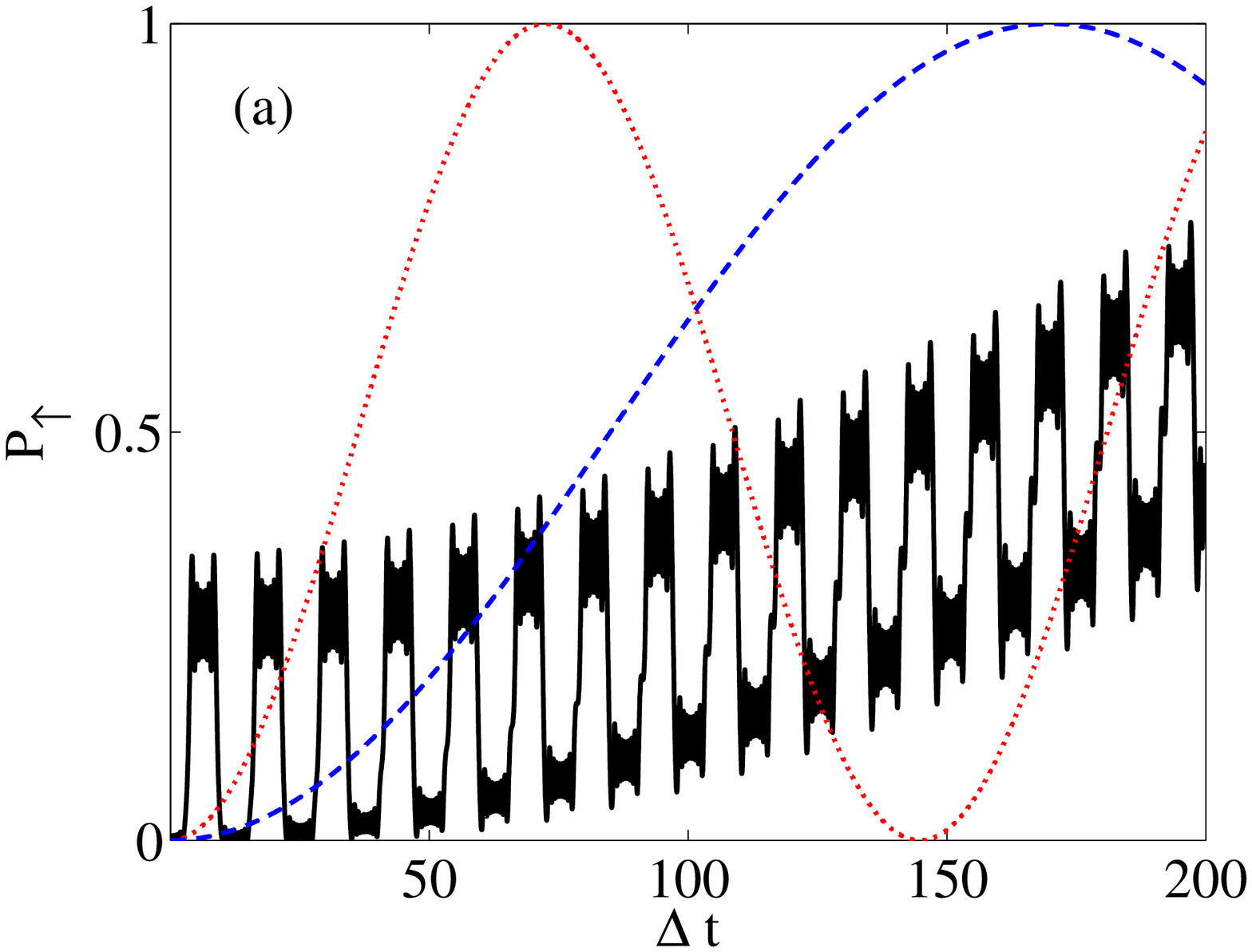}
\includegraphics[width=6.5cm]{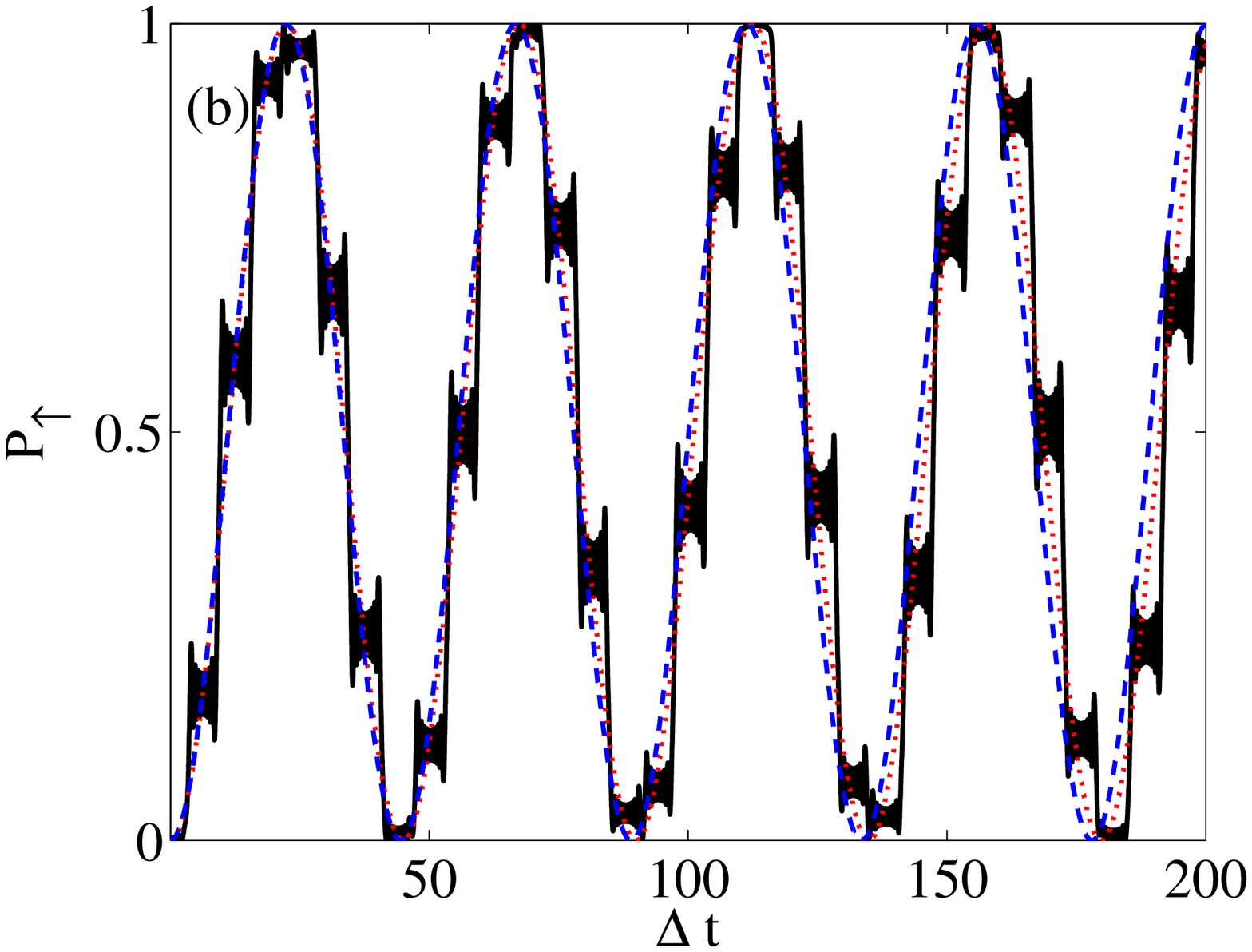}
\includegraphics[width=6.5cm]{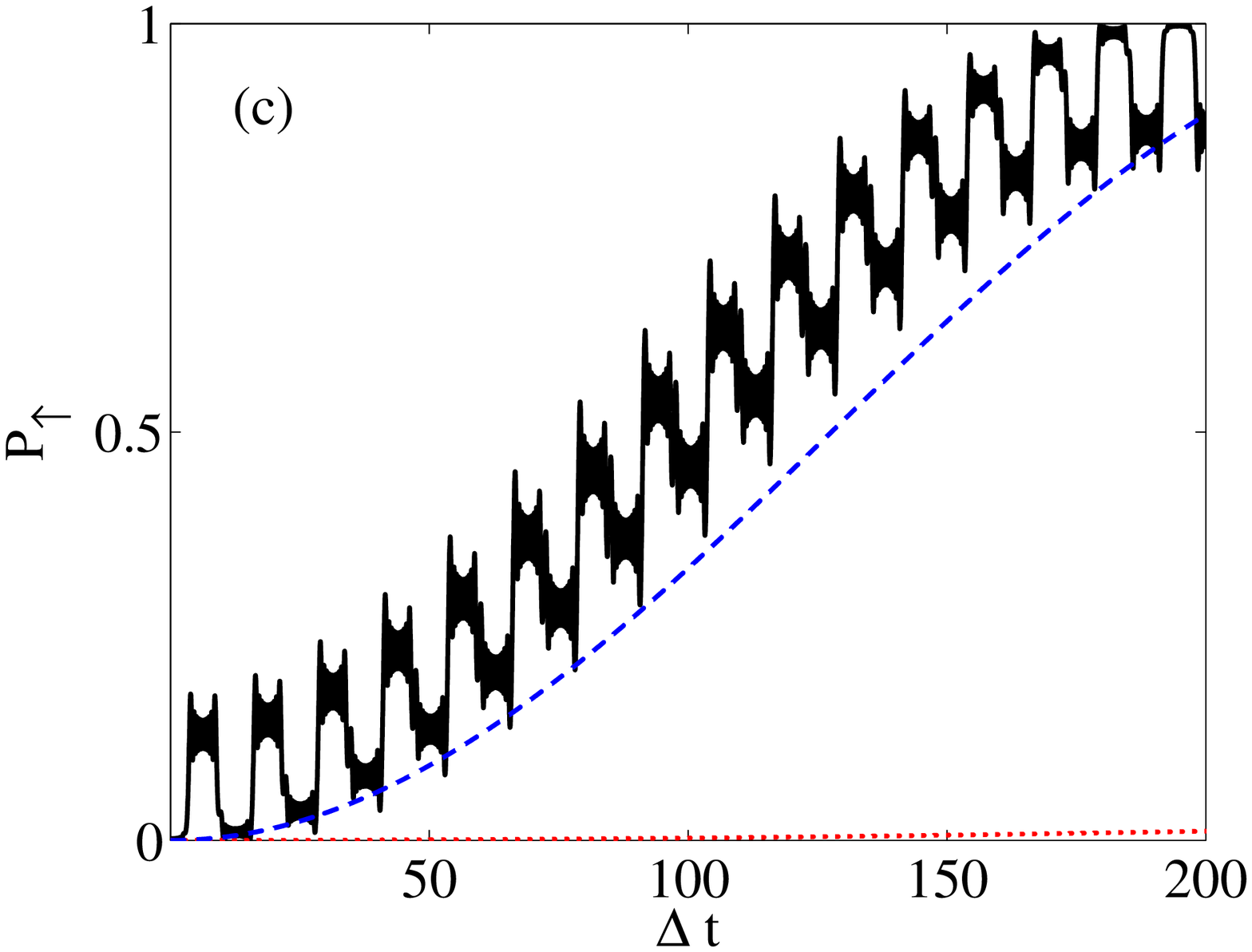}
\caption{(color online) Occupation probability $P_{\uparrow}$ of
the state $\left|\uparrow\right\rangle$ as a function of time (in
dimensionless units), assuming that the system was initially in
the state $\left|\downarrow\right\rangle$. In all the figures,
$\epsilon_0/\Delta=1$, and $\omega/\Delta=0.5$. The driving
amplitude is given by $A/\Delta=$12(a), 16(b) and 20(c). The blue
dashed line gives the (coarse-grained) predicted dynamics from the
transfer-matrix method in the large-amplitude limit, and the red
dotted line gives the predicted dynamics from Sec.~III.}
\end{figure}

The fact that the resonance condition is always satisfied in the
unbiased case is very clear in this approach. In this case, the
phase factors $\theta_1$ and $\theta_2$ accumulated on the two
sides cancel because of symmetry, regardless of the driving
amplitude and frequency.

The factor $|\cos(\tilde{\theta}_2-\pi/4)|$ in Eq.
(\ref{eq:TMOscillationFrequency}) gives a nontrivial dependence of
the $\left|\uparrow\right\rangle\leftrightarrow
\left|\downarrow\right\rangle$ oscillation frequency on the bias
and driving parameters. It lies behind the phenomenon that even if
the resonance condition is satisfied, it is still possible for the
oscillations to be so slow that the resonance is effectively
destroyed (see Fig. 3). In other words, it describes the same
mechanism responsible for CDT, and it agrees with the Bessel
ladder structure discussed in Ref.~\cite{Oliver}.

\subsection{Slow-crossing limit}

When $\Delta^2/\omega\sqrt{A^2-\epsilon_0^2} \gg 1$, we find that
\begin{equation}
\cos^2 \displaystyle\frac{\chi}{2} \approx \exp \left\{ -
\frac{\pi\Delta^2}{2v} \right\},
\end{equation}
and $\sin^2(\chi/2) \gg \cos^2(\chi/2)$.

The angle $\theta_{\rm FC}$ is now given by
\begin{eqnarray}
\theta_{\rm FC} & \approx & - 2 (\theta_{\rm LZ, 1} - \theta_{\rm
LZ, 2} + \theta_1 - \theta_2 )
\nonumber
\\
& \approx & - 2 \pi + \frac{2\pi\epsilon_0}{\omega} + 4
\frac{\sqrt{A^2-\epsilon_0^2}}{\omega} - 4
\frac{\epsilon_0}{\omega} \cos^{-1} \frac{\epsilon_0}{A}
\nonumber
\\
& & + 2 f_1 + 2 f_2,
\label{eq:thetaFC_slow_limit}
\end{eqnarray}
where $f_1$ and $f_2$ grow, roughly speaking, as
$(\Delta^2/A\omega)\log(A/\Delta)$ (see Eqs.~\ref{eq:LZphases} and
\ref{eq:theta_tildes}). Because there are no simple expressions
for $f_1$ and $f_2$ (combined with the fact that they are much
larger than unity in the slow-crossing limit), deriving an
analytic expression for the resonance condition is not as
straightforward in this case as in the fast-crossing limit. The
resonance lines can be obtained numerically, since a numerical
calculation of $\tilde{\theta}_1$ and $\tilde{\theta}_2$ would be
straightforward. Alternatively, one can obtain a rough idea about
the shapes of the resonance lines by ignoring $f_1$ and $f_2$ in
comparison to the other terms in
Eq.~(\ref{eq:thetaFC_slow_limit}). Equating $\theta_{\rm FC}$ to
$2\pi n$, the rough approximation of the resonance condition with
the above simplification is given by
\begin{equation}
\frac{\epsilon_0}{\omega} + 2 \frac{\sqrt{A^2-\epsilon_0^2}}{\pi
\omega} - 2 \frac{\epsilon_0}{\pi \omega} \cos^{-1}
\frac{\epsilon_0}{A} = n.
\end{equation}
In the $A$-$\epsilon_0$ plane, the resonance lines form arcs
around the origin (i.e. around the point $A=\epsilon_0=0$). Note
that when $\epsilon_0=0$, the above resonance condition reduces to
\begin{equation}
\frac{2A}{\pi \omega} = n,
\end{equation}
such that the parameter $A$ is crucial in the resonance condition,
in contrast to the results of the fast-crossing limit. The
oscillation frequency can also be calculated numerically using
Eq.~(\ref{eq:TMOscillationFrequencyDerivation}). The angle
$\phi_{\rm FC}$ is given by
\begin{equation}
\phi_{\rm FC} \approx \tilde{\theta}_1 + \tilde{\theta}_2.
\end{equation}

\section{Discussion and conclusion}

\begin{figure}[h]
\includegraphics[width=7.0cm]{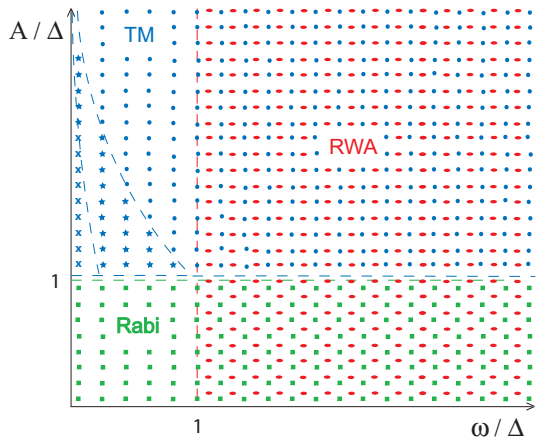}
\caption{(color online) Regions of validity for different
approximations. TM stands for transfer-matrix method, RWA stands
for the rotating-wave approximation presented in this paper, and
Rabi stands for the weak-driving limit best-known in connection
with Rabi oscillations. The axes are the frequency $\omega$ and
amplitude $A$ of the driving field, both normalized to the minimum
gap $\Delta$. The Rabi region is described by the condition
$A/\Delta<1$ (regardless of the value of $\omega$), and it is
shown by green squares. The RWA region is described by the
condition $\omega/\Delta>1$, and it is shown by red ellipses. The
TM region is described by the conditions $A/\Delta>1$, and it is
shown by blue symbols. The TM region can be divided into three
sub-regions depending on the parameter $A\omega/\Delta^2$: the
slow-crossing limit (x symbols), the fast-crossing limit (circles)
and the intermediate-speed case (stars). Note that $\epsilon_0$
was generally assumed to be comparable to $\Delta$ in this
figure.}
\end{figure}

In this paper we have presented two approaches to study the
problem of a strongly driven two-level system. The first one
(presented in Sec.~III) was based on a rotating-wave
approximation, whereas the second one (presented in Sec.~IV) was
based on a discretized description of the dynamics. It is
important to note that the two approaches have different regions
of validity, as shown in Fig.~6. We also include in Fig.~6 the
region of validity for the weak-driving limit, which is most
clearly associated with Rabi-oscillation physics. As can also be
seen in Fig.~6, the different regions overlap substantially, and
two of them can sometimes be used to describe the same situation.
For example, although the derivation and the appearance of the
results of Sec.~IV rely strongly on the picture of LZ crossings,
the results agree with those of Sec.~III whenever
$\omega\gg\Delta$. Naturally, the deeper one goes into one of
these regions, the better the results one can expect to obtain
from the corresponding approximation.

The TM region is divided into three sub-regions. When $A\omega \gg
\Delta^2$, each LZ crossing occurs in the fast limit and the
expressions given in Sec.~IV.A are valid. In the opposite limit,
i.e.~when $A\omega \ll \Delta^2$, the crossings occur in the slow
limit, and Sec.~IV.B applies. Between these two limits, one has
the intermediate-speed, or general, case. Even though we have not
derived any quantitative results describing the dynamics in this
case, it can be conceptually understood using the TM picture
discussed here (as can be seen from Fig.~5).

The approaches discussed here therefore cover a large portion of
the parameter space. They provide alternative points of view for
understanding the mechanisms at play in the dynamics of this
system \cite{Rotvig}. It is worth noting that the approach of
Sec.~IV is not limited to harmonic driving. It can be used to
treat any system with large-amplitude driving around the
degeneracy point, assuming the approximation of linear sweeps
through the crossing region is valid \cite{Peeters}.

Experiments on two-level systems have generally suffered from
short coherence times. With the advent of the field of QIP, the
need for long coherence times has spurred a fast advance in the
direction of isolating qubits from their environments, thus
resulting in relatively long coherence times. For example,
high-order processes and quantum interference between LZ crossings
have already been observed in superconducting qubit systems
\cite{Oliver,Sillanpaa,SCexperiments,You}. One could in the future
realistically think about using strong driving as a tool to
manipulate qubits. The mechanisms discussed in this paper can be
used in constructing such qubit-manipulation tools.

\begin{acknowledgments}
We would like to thank S. Shevchenko for valuable comments. This
work was supported in part by the National Security Agency (NSA),
the Army Research Office (ARO), the Laboratory for Physical
Sciences (LPS), the National Science Foundation (NSF) grant
No.~EIA-0130383 and the JSPS CTC program. One of us (S.A.) was
supported by the Japan Society for the Promotion of Science
(JSPS).
\end{acknowledgments}


\begin{thebibliography}{99}
\bibitem{Basis} Here we treat the $\hat{\sigma}_z$ term as the
externally tunable term, and we use the eigenstates of
$\hat{\sigma}_z$, to which we refer as
$\left|\uparrow\right\rangle$ and $\left|\downarrow\right\rangle$,
as the preferred basis in our analysis.
\bibitem{DroppingBessel} Note that, in order to achieve simpler
appearance, we have dropped the Bessel functions from the
right-hand side of the inequality, because they are generally of
order one. This explains why we shall find cases where
$\omega<\Delta$, but the RWA still gives good agreement with the
exact results.
\bibitem{AsymptoticFrequency} From Eq.~(\ref{eq:asymptotics}),
with $z\gg n$, we can see that the frequency of
$\left|\uparrow\right\rangle \leftrightarrow
\left|\downarrow\right\rangle$ oscillations scales as
$\sqrt{\omega/A}$, apart from the cosine factor. The larger the
driving amplitude (within a certain region), the slower the
oscillations, in agreement with our intuitive guess.
\bibitem{SlowDynamics} The $\left|\uparrow\right\rangle \leftrightarrow
\left|\downarrow\right\rangle$ oscillation dynamics occurs on a
much longer time scale than $1/\omega$, such that the exact
details of the boundary conditions do not have a significant
impact on the dynamics.
\bibitem{GeometricArgument} In fact, starting from
Eq.~(\ref{eq:DiscretePropagatorSeparated}) one can use
straighforward geometric arguments to show that the evolution of
the system over a full driving cycle in the non-resonant case can
be represented by a single rotation about an axis that is almost
parallel to the $z$-axis. This means that we cannot obtain
$\left|\uparrow\right\rangle\leftrightarrow
\left|\downarrow\right\rangle$ oscillations by repeated
application of this rotation.
\bibitem{PlottedTM} Note that the TM approach can also be used to
obtain the step-like dynamics of this system. However, here we
only show `coarse-grained' sinusoidally oscillating functions for
clarity.
\end{thebibliography}
\end{document}